\documentclass[pre,showpacs,superscriptaddress,twocolumn]{revtex4}

\usepackage{graphicx}
\usepackage{amsmath}
\usepackage{amssymb}
\usepackage{mathrsfs}

\usepackage[british]{babel}

\begin{document}

\title{Anomalous diffusion in correlated continuous time random walks}

\author{Vincent Tejedor}
\affiliation{Physics Department T30g, Technical University of Munich,
85747 Garching, Germany}
\author{Ralf Metzler}
\affiliation{Physics Department T30g, Technical University of Munich,
85747 Garching, Germany}

\begin{abstract}
We demonstrate that continuous time random walks in which successive waiting
times are correlated by Gaussian statistics lead to anomalous diffusion with
mean squared displacement
$\langle\mathbf{r}^2(t)\rangle\simeq t^{2/3}$. Long-ranged
correlations of the waiting times with power-law exponent $\alpha$
($0<\alpha\le2$) give rise to subdiffusion of the form $\langle\mathbf{r}^2(t)
\rangle\simeq t^{\alpha/(1+\alpha)}$. In contrast correlations in the jump
lengths are shown to produce superdiffusion. We show that in both cases weak
ergodicity breaking occurs. Our results are in excellent agreement with
simulations.
\end{abstract}

\pacs{02.50.Ey,05.40.-a}

\maketitle

\section{Introduction}

The continuous time random walk (CTRW) theory was introduced more than forty
years ago \cite{Montroll} to extend regular random walks on lattices to a
continuous time variable. We characterise each jump in a CTRW process by a
waiting time $\tau$ elapsing after the previous jump, and a variable jump
length $\xi$. Each $\tau$ and $\xi$ are independent random variables,
identically distributed according to the probability densities $\psi(\tau)$
and $\lambda(\xi)$. For a power-law form $\psi(\tau)\simeq\tau^{-1-\beta}$
with $0<\beta<1$ the characteristic waiting time $\int_0^{\infty}\tau\psi(
\tau)d\tau$ diverges and the resulting CTRW is subdiffusive, $\langle\mathbf{r}
^2(t)\rangle\simeq t^{\beta}$. For $\lambda(\xi)\simeq|\xi|^{-1-\gamma}$ with
$0<\gamma<2$ the jump length variance $\int_{-\infty}^{\infty}\xi^2\lambda
(\xi)d\xi$ diverges and we obtain a superdiffusive L{\'e}vy flight.
Spatiotemporal coupling of jump length and waiting time leads to L{\'e}vy
walks with finite $\langle\mathbf{r}^2(t)\rangle\simeq t^{\beta}$
with $1<\beta<2$ \cite{klablushle}.
The CTRW model was championed in the seminal work
on charge carrier transport in amorphous semiconductors \cite{scher}.
CTRW theory has also been successfully applied
in subsurface tracer dispersion \cite{geol}, tick-tick dynamics in financial
markets \cite{finance}, cardiological rhythms \cite{cardiology}, electron
transfer \cite{electron},
noise in plasma devices \cite{chechkin}, dispersion in turbulent systems
\cite{grigolini}, search models \cite{lomholt}, or in models of gene
regulation \cite{lomholt1}, among many others \cite{report}. CTRW models are
closely related to the fractional Fokker-Planck equation \cite{ffpe,report}.

The independence of the waiting times and jump lengths giving rise to a renewal
process is not always justified. As soon as the random walker has some form of
memory, even a short one, the variables become non-independent. Examples are
found in financial market dynamics \cite{finance_corr}, single trajectories
in which there is a directional memory \cite{speed_corr}, or in astrophysics
\cite{astrology_corr}. An important application are search processes and human
motion patterns in which memory and conscience will likely lead to a
non-renewal situation. A general mathematical framework was developed for
non-independent CTRWs \cite{Correlated}, however, it is quite cumbersome to
apply and the cases solved so far only lead back to normal diffusion.
An approach to coupling of waiting times based on the Langevin equation
formulation of CTRW processes was recently introduced \cite{igor}. Some
special cases were explored that bridge between CTRW and fractional Brownian
motion \cite{meershaert}. Here we introduce a simple way to establish
correlations in CTRW processes. Correlations between successive waiting times
are shown to give rise to subdiffusion even when they are Gaussian, while
correlations between jump lengths produce superdiffusion. We also consider
long-ranged correlations and discuss anticorrelation effects.
Our scaling arguments are in excellent accord with simulations.

After introducing the general framework we demonstrate how correlated waiting
times lead to subdiffusion and weak ergodicity breaking. We then consider
correlations in the jump length, and proceed to analyse anticorrelation
effects. Finally we provide some details on the derivations and draw our
conclusions.

\section{General framework}

Let us briefly review the general framework for correlated CTRWs
from Ref.~\cite{Correlated} for correlated
waiting times. Then the waiting time $\psi_n$ for a given step $n$ is
conditioned by the previous waiting time $\psi_{n-1}$, as quantified
by the joint probability $P(\psi_n,\psi_{n-1})=P(\psi_n|\psi_{n-1})
P(\psi_{n-1})$. Here $P(\psi_{n-1})$ is the probability to have the
waiting time $\psi_{n-1}$ in the previous step, and $P(\psi_n|\psi_{n-1})$
is the conditional probability to find a waiting time $\psi_n$ for given
$\psi_{n-1}$. The normalisation conditions are
\begin{eqnarray}
\nonumber
&&\int_{0}^{\infty}P(\psi_n)d\psi_n=1,\,\,\,\int_{0}^{\infty}\int_{0}^{\infty}
P(\psi_n,\psi_{n-1})d\psi_nd\psi_{n-1}=1,\\
&&\int_{0}^{\infty}P(\psi_n|\psi_{n-1})d\psi_n=1.
\label{correl}
\end{eqnarray}
By recurrence, we obtain the joint probability
\begin{equation}
\label{joint}
P(\psi_n,\psi_{n-1},\ldots,\psi_0)=\prod_{k=1}^nP(\psi_k|\psi_{k-1})P(\psi_0),
\end{equation}
demonstrating that the waiting time $\psi_n$ in fact depends on all previous
waiting times. Note that in the decoupled case $P(\psi_k|\psi_{k-1})=
f(\psi_k)g(\psi_{k-1})$ we get back to a regular renewal CTRW.

The marginal probability of $\psi_n$ is defined as
$P(\psi_n)=\int_0^{\infty}P(\psi_n|\psi_{n-1})P(\psi_{n-1})d\psi_{n-1}$.
According to Eq.~(\ref{joint}), this leads to an $n$-fold integration over
the product on its right hand side.
In the general case, it is quite hard to compute this quantity, and this is
why in previous literature only normal diffusion was treated in this
framework.

\section{Random walk of waiting times}

Instead of constructing
the process from definitions (\ref{correl}) and (\ref{joint}) we start from a
different angle. Namely we build the value of waiting time $\psi_n$ from the
previous waiting time plus a random deviation $\delta\psi_n$,
\begin{equation}
\label{delta}
\psi_n=\psi_{n-1}+\delta\psi_{n}.
\end{equation}
The sequence of waiting times can therefore be viewed as a random walk in the
space of waiting times. Similarly we will proceed with coupled jump lengths
below.
The waiting time $\psi_n$ can therefore be expressed through
$\psi_n=\sum_{i=0}^{n}\delta\psi_i$,
where we assumed that $\psi_0=\delta\psi_0$, without loss of generality. A
reflecting boundary condition at $\psi =0$ ensures that the waiting times are
positive. If the random variations $\delta\psi_n$ are normally distributed,
we obtain the following conditional probability:
\begin{eqnarray}
\nonumber
P(\psi_n|\psi_{n-1})&=&\frac{1}{\sqrt{4\pi\sigma^2}}\left[\exp\left(-\frac{(
\psi_n-\psi_{n-1})^2}{4\sigma^2}\right)\right.\\
&&+\left.\exp\left(-\frac{(\psi_n+\psi_{n-1})^2}{4\sigma^2}\right)\right].
\label{Ppsi}
\end{eqnarray}
The mean squared displacement (MSD) for this process grows with the number of
steps as $\langle(\Delta\psi(n))^2\rangle\sim 2\sigma^2 n$ for large $n$.
To proceed, we compute
the probability $P(t,n)$ to have made $n$ steps up to time $t$. In Laplace
space \cite{REM}
\begin{equation}
P(s,n)=\int_0^{\infty}P(t,n) e^{-st}dt=\frac{1-\psi_n(s)}{s}\prod_{i=0}^{n-1}
\psi_i(s).
\label{Psn}
\end{equation}
Here and in the following we use the convention that the Laplace transform
of a function $f(t)$ is expressed by explicit dependence on the Laplace
variable, that is, $\mathscr{L}\{f(t)\}=f(s)$.
After some calculations we arrive at (see Appendix \ref{appendix})
\begin{eqnarray}
\nonumber
P(s,n)&=&\frac{1}{s}\left[\delta\psi\left(\sqrt{\frac{n(n+1)(2n+1)}{6}}s\right)
\right.\\
&&\left.-\delta \psi\left(\sqrt{\frac{(n+1)(n+2)(2n+3)}{6}}s\right)\right].
\label{Psn2}
\end{eqnarray}
We obtain the Laplace transform of the mean number of steps by summation,
$\langle n(s)\rangle=\sum_{n=0}^{\infty}nP(s,n)$. With the approximations
detailed in Appendix \ref{appendix}, we find in leading order around $s=0$
(corresponding to long times),
\begin{equation}
\label{number}
\langle n(s)\rangle\sim\frac{1}{s}\frac{3^{1/3}\Gamma(5/3)}{(s\tau)^{2/3}}
\Rightarrow\langle n(t)\rangle\sim\left(\frac{t}{\sigma}\right)^{2/3}
\end{equation}
where we assumed $\tau\propto\sigma$. At long times the Gaussian waiting
time correlation results in the subdiffusion law \cite{hughes}
\begin{equation}
\label{msd}
\langle\mathbf{r}^2(t)\rangle=\langle\delta\mathbf{r}^2\rangle\langle n(t)
\rangle\sim Kt^{2/3},
\end{equation}
where $\langle\delta\mathbf{r}^2\rangle$ is the jump length variance of the
process and $K=\langle\delta\mathbf{r}^2\rangle/\sigma^{2/3}$ the generalised
diffusion constant. Eq.~(\ref{msd}) is one of the main result of this work.
We stress again that this subdiffusion emerges from a random process that has
a finite waiting time in each step. However, as time proceeds this waiting time
slowly diverges [$\langle\Delta\psi(n)\rangle\sim n^{1/2}$], due to the
diffusive coupling of waiting times.

If we take a sharp correlation of waiting times with $\sigma=0$, 
Eq.~(\ref{Ppsi}) leads to $P(\psi_n)=\delta(\psi_n-\psi_0)+\delta(\psi_n+
\psi_0)$. At each step the waiting time is the same. The mean waiting time
is $\langle \psi_n\rangle=\int_0^{\infty}\psi\left[\delta(\psi-\psi_0)+\delta(
\psi+\psi_0)\right]d\psi=\psi_0$ such that we find classical Brownian motion
with diffusion coefficient $\langle\delta\mathbf{r}^2\rangle/\psi_0$, as it
should be.

Consider now the case of a stable distribution with index $\alpha$ for $\delta
\psi$. For $1<\alpha\le2$ the first moment $\langle\delta\psi\rangle=\tau$ of
the random walk in $\psi$ space is still finite. We then follow similar
steps as outlined above, obtaining
\begin{equation}
\label{nscaling}
\langle n(t)\rangle\sim\left(t/\sigma\right)^{\alpha/(\alpha+1)}\,\Rightarrow
\,\langle\mathbf{r}^2(t)\rangle\simeq Kt^{\alpha/(\alpha+1)}.
\end{equation}
The case $0<\alpha\le 1$ is somewhat more involved. We argue that for any
stable distribution, we have $\sum_{i=1}^n\psi_i(t){\buildrel d\over\sim}
n^{-(\alpha+1)/\alpha}\delta\psi(n^{-(\alpha+1)/\alpha}t)$, where ${\buildrel
d\over\sim}$ is a scaling equality of distributions. This leads to the relation
$t(n)/\sigma{\buildrel d\over\sim}n^{(\alpha+1)/\alpha}$, such that
Eq.~(\ref{nscaling}) still holds for any $0<\alpha\leq2$. We note that
result (\ref{nscaling}) was retrieved from a Langevin equation approach in
Ref.~\cite{igor}.

\begin{figure}
\includegraphics[height=7.2cm,angle=270]{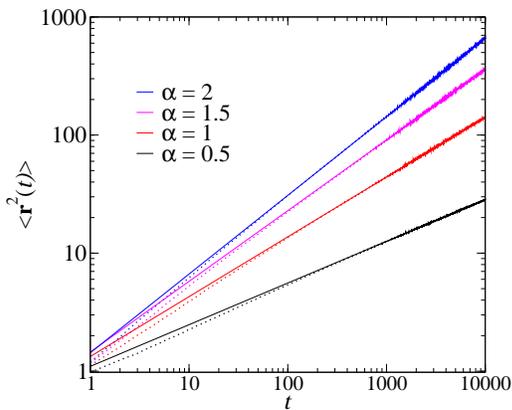} 
\caption{$\langle\mathbf{r}^2(t)\rangle$ for a waiting time correlated 3D
Gaussian walk. The $\delta\psi$ follow an $\alpha$-stable law with
scale factor $c=1$; $\alpha$ decreases from top to
bottom. Simulations (---) and power-laws ($\cdots$) with fitted
exponents
$0.35,0.50,0.60,0.66$. Theoretical values $\alpha/(\alpha+1)$: $0.33,0.50,
0.60,0.66$.}
\label{levy}
\end{figure}

For a stable distribution the jumps between successive waiting times
become increasingly larger when $\alpha$ is decreased, effecting even slower
diffusion $\langle\mathbf{r}^2(t)\rangle\simeq t^{\alpha/(\alpha+1)}$. In the
limit $\alpha=2$ we are back to Gaussian diffusion in $\psi$ space and a $2/3$
subdiffusion in position space, $\langle\mathbf{r}^2(t)\rangle\simeq t^{2/3}$.
In Fig.~\ref{levy} we demonstrate excellent agreement between our analytical
findings and simulations results of random
processes performed with stable correlations between successive waiting times.

\begin{figure}
\includegraphics[height=7.2cm,angle=270]{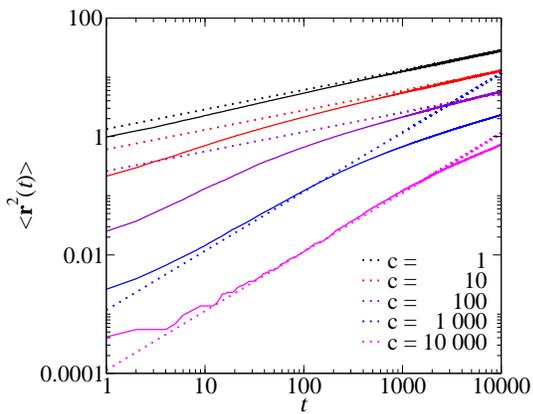} 
\caption{Same as Fig.~\ref{levy} with $\alpha=0.5$ and various scale factors
$c$ ($c$ increases from top to bottom). Simulations (---) and power-laws
($\cdot$) with slope $0.33$ ($c=1,10,100$) and $1$ ($c=1.000,10.000$).
Note the turnover to slope $\alpha/(\alpha+1)$ at $t \sim c$. Compare text.}
\label{levyc}
\end{figure}

The characteristic function of the stable variable $\delta\psi$ is $\phi_{
\delta\psi}(q)=\exp\left(-|cq|^{\alpha}\right)$, see Refs.~\cite{hughes}.
While at long times the
predicted subdiffusive behaviour is attained, we observe a transient regime of
normal diffusion, $\langle\mathbf{r}^2(t)\rangle\simeq t$, when the scale
factor $c$ increases (see Fig.~\ref{levyc}). At short times, we may neglect
the probability that the random walker makes more than one step. The
probability to make the first step corresponds to the cumulative function of
the waiting time distribution $F_{\delta \psi}(t)$,
\begin{equation}
F_{\delta\psi}(t)=\frac{2}{\pi}\int_{0}^{\infty}\frac{\sin(tq)}{q}\phi_{
\delta\psi}(q)dq\sim\frac{2\Gamma(1+1/\alpha)}{\pi c}t.
\end{equation}
Thus, the initial linear slope in the ensemble average is due to the linearity
of the cumulative function for short waiting times. This effect vanishes as
soon as the probability increases that the walker makes two or more steps, and
converges to the predicted long time behaviour (\ref{nscaling}).

\subsection{Weak ergodicity breaking}

\begin{figure}
\includegraphics[height=8.8cm,angle=270]{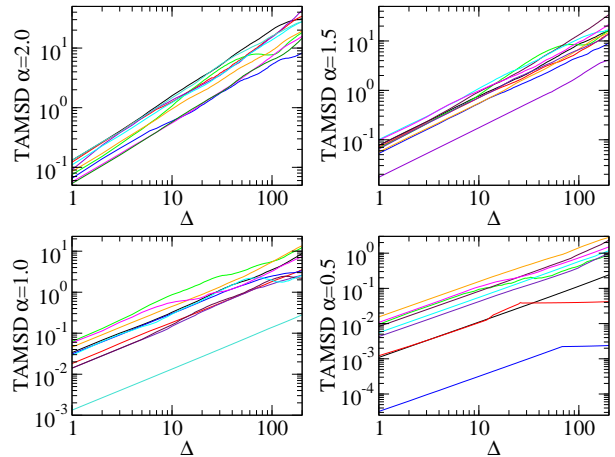}
\caption{Time-averaged MSD as function of $\Delta$ for a CTRW with correlated
waiting time ($T=2000$). The plots show a scatter between individual
trajectories that becomes more pronounced for decreasing $\alpha$. The data
approximately scale as the ensemble average, Eq.~(\ref{tadelta}) .
For very small $\alpha$ we observe a plateau due to the occurrence of
extremely long waiting times of the order of $T$.}
\label{scatter}
\end{figure}

When dealing with the time series of a single particle trajectory of length
$T$, instead of the ensemble averaged MSD $\langle\mathbf{r}^2(t)\rangle$ one
calculates the time averaged MSD
\begin{equation}
\label{tamsd}
\overline{\delta^2(\Delta,T)}=\frac{1}{T-\Delta}\int_0^{T-\Delta}\Big[
\mathbf{r}(t+\Delta)-\mathbf{r}(t)\Big]^2dt,
\end{equation}
relating two positions separated by the lag time $\Delta$. Ensemble averaging
Eq.~(\ref{tamsd}) the square brackets become $\langle\left[\mathbf{r}(t+\Delta)
-\mathbf{r}(t)\right]^2\rangle=\langle\delta\mathbf{r}^2\rangle\langle n_{t,t+
\Delta}\rangle$ where $\langle\delta\mathbf{r}^2\rangle$ is the (finite)
variance of jump lengths, and $\langle n_{t,t+\Delta}\rangle$ counts the
average number of jumps in the time interval $[t,t+\Delta]$. For normal
diffusion $\langle n(t)\rangle=t/\tau$ and therefore $\overline{\delta^2(
\Delta,T)}=2dK\Delta$ behaves exactly as the ensemble average such that the
lag time $\Delta$ is exchangeable with the process time $t$ of the ensemble
average. In contrast for a subdiffusive renewal CTRW with waiting time density
$\psi(t)\simeq t^{-1-\alpha}$ ($0<\alpha<1$) it turns out that $\overline{
\delta^2(\Delta,T)}\simeq K\Delta/T^{1-\alpha}$, i.e., we observe a so-called
weak ergodicity breaking for $\Delta\ll T$ \cite{web,web1}. Do we observe
similar
behaviour for our coupled CTRW? With Eq.~(\ref{number}) and the relation
$\langle n_{t,t+\Delta}\rangle=\langle n_{0,t+\Delta}\rangle-\langle n_{0,t}
\rangle$,
\begin{equation}
\left<\big[\mathbf{r}(t+\Delta)-\mathbf{r}(t)\big]^2\right>\sim 2K
\left[(t+\Delta)^{\alpha/(1+\alpha)}-t^{\alpha/(1+\alpha)}\right],
\end{equation}
with $K=\langle\delta\mathbf{r}^2\rangle/(2d\sigma^{\alpha/(1+\alpha)})$,
and therefore
\begin{equation}
\label{tadelta}
\left<\overline{\delta^2(\Delta,T)}\right>\sim 2dK\Delta/T^{\alpha/(1+\alpha)}.
\end{equation}
Thus our non-renewal, coupled CTRW also exhibits a weak ergodicity breaking,
even in the limit $\alpha=2$ of Gaussian waiting time coupling. Simulations
of this process indeed confirm the predicted scaling of the time averaged
MSD with both lag time $\Delta$ and measurement time $T$. In Fig.~\ref{scatter}
we demonstrate that individual trajectories show significant scatter in their
amplitude, as observed for renewal CTRW subdiffusion \cite{web}.

\section{Correlated jump lengths}

\begin{figure}
\includegraphics[height=7.2cm,angle=270]{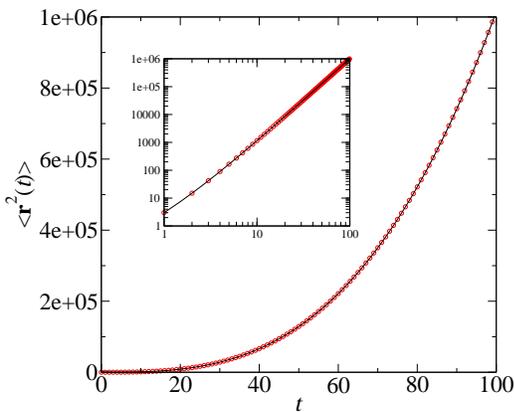} 
\caption{3D random walk with correlated jump lengths. We chose
$\mathbf{x}(0)=0$. $\delta\mathbf{x}(t)$ is normally
distributed with $\sigma=\sqrt{2}$. Simulations ($\circ$) and theory (---),
Eq.~(\ref{eq:msdjump}).}
\label{jump}
\end{figure}

Let us now consider a random walk with constant waiting time $\psi=1$ but
correlated jump lengths. If $\mathbf{r}(t)$ is the position of the walker
we define $\mathbf{x}(t)=\mathbf{r}(t)-\mathbf{r}(t-1)$ ($\mathbf{x}(0)=
0$). The jump length is now assumed to diffuse in $\mathbf{x}$ space,
with increments $\delta\mathbf{x}(t)=\mathbf{x}(t)-\mathbf{x}(t-1)$ that are
normally distributed with mean $0$ and variance $\sigma$. We find
\begin{equation}
\label{gaussjump}
P\Big(\mathbf{x}(t)=(x,y,z)\Big)=\frac{1}{(\pi\sigma^2t)^{3/2}}\exp\left(
-\frac{x^2+y^2+z^2}{\sigma^2 t}\right)
\end{equation}
with variance $\langle\mathbf{x}^2(t)\rangle=\frac{3}{2}\sigma^2t$. For this
process we have $\mathbf{r}(t)=\mathbf{r}(0)+\sum_{i=1}^{t}\mathbf{x}(i)$
and obtain (see Appendix \ref{appendix})
\begin{equation}
\langle[\mathbf{r}(t)-\mathbf{r}(0)]^2\rangle=\frac{t(t+1)(2t+1)\sigma^2}
{4}.
\label{eq:msdjump}
\end{equation}
Thus, when we assume a diffusion in the space of jump lengths we obtain a
Richardson-type $t^3$ scaling. Fig.~\ref{jump} demonstrates excellent
agreement of Eq.~(\ref{eq:msdjump}) with the simulations result.
For a fixed jump length ($\sigma=0$) all steps have the same length and
direction, and the corresponding random walk is ballistic, $\langle\mathbf{r}
^2(t)\rangle=\mathbf{x}^2(0)t^2$.

Classical diffusion cannot be retrieved with this mechanism, as this would
require directional randomness for jumping left/right. In fact we obtain for
the jump correlations that $\langle \mathbf{x}(t)\cdot\mathbf{x}(t+\tau)\rangle
=\frac{3}{2}\sigma^2t$. The position correlations become
\begin{equation}
\label{autocorr}
\langle\mathbf{r}(t)\cdot\mathbf{r}(t+\tau)\rangle=\frac{t(t+1)(2t+1)\sigma
^2}{4}+\frac{3\sigma^2 t(t+1)\tau}{4}.
\end{equation}

If the increments $\delta \mathbf{x}$ are distributed according to a stable 
distribution of exponent $\alpha$ and scale factor $c$, each coordinate of
$\mathbf{r}(t)$ is distributed according to a stable distribution of
exponent $\alpha$ and scale factor $c(t)=\left(\sum_{i=1}^ti^{\alpha}\right)
^{1/\alpha}c\simeq ct^{(1+\alpha)/\alpha}$. The mean squared displacement
diverges
for $0<\alpha<2$ but we observe the superdiffusive scaling $x\sim t^{(1+\alpha)
/\alpha}$. In the limit $\alpha=2$, using $2c^2=\sigma^2$
we retrieve the previous result (\ref{eq:msdjump}), while for $\alpha\to0$
we find regular L{\'e}vy flight scaling $x\sim t^{1/\alpha}$ \cite{report}.

\subsection{Weak ergodicity breaking}

\begin{figure}
\includegraphics[height=8.0cm,angle=270]{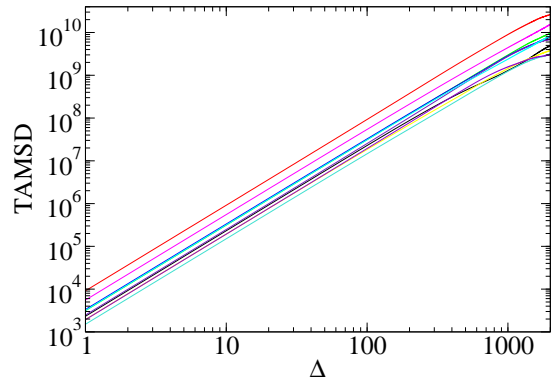}
\caption{Time-averaged MSD as function of $\Delta$ for ten different
trajectories of length $T=2000$ with jump length correlation. We chose
$\sigma^2=2$.}
\label{scatter1}
\end{figure}

The process with coupled jump lengths has the same waiting time for each
jump. Could it still be subject to weak ergodicity breaking? Combining
Eqs.~(\ref{tamsd}) and (\ref{autocorr}) we obtain the equality
\begin{equation}
\left<\overline{\delta^2}(\Delta,T)\right>=\frac{3\sigma^2}{4}\Delta^2 T+
\sigma^2\left(\frac{\Delta}{4}+\frac{3\Delta^2}{4}-\frac{\Delta^3}{4}\right).
\end{equation}
In the limit $\Delta\ll T$ we find the scaling $\left<\overline{\delta^2}(
\Delta,T)\right>\simeq\Delta^2 T$, contrasting the leading cubic behaviour
$\langle\mathbf{r}^2(t)\rangle\simeq t^3$ of the ensemble average. Thus,
also the non-renewal CTRW with coupled jump lengths leads to a weak ergodicity
breaking. Again, simulations corroborate the scaling with lag time $\Delta$
and overall process time $T$. In Fig.~\ref{scatter1} we show the scatter
between different trajectories.

\section{Anticorrelated waiting times}

Another way to introduce correlations is to consider diffusion on a network
of nodes on each of which a different distribution is assumed.
We illustrate this approach by anticorrelated waiting times. Namely
we have two waiting time distributions, $\Psi_1$ and $\Psi_2$, such that
$\Psi_1$ is centred around short waiting times and $\Psi_2$ around longer
ones. We start by choosing a waiting time from one of the $\Psi_i$
and then pass from one waiting time distribution to the other according to the
followings rules: (i) if we are in state $\Psi_1$ and get a waiting time
shorter than a preset $t_1$, we shift to $\Psi_2$ for the next step. Otherwise
we remain with $\Psi_1$. (ii) If we are in state $\Psi_2$ and find a
waiting time longer than a given $t_2$, we shift to $\Psi_1$, otherwise stay
in $\Psi_2$.

We can view this process as a diffusion in a network with two nodes. Being at
node $i$ at step $n$ means that the $n$th waiting time is
extracted from distribution $\Psi_i$. The transition probabilities
between the two nodes are $P_1=\int_{0}^{t_1}\Psi_1(t)dt$ and $P_2=\int_{t_2}
^{\infty}\Psi_2(t)dt$. The probability that the $n$th waiting time
is chosen from distribution $\Psi_1$ is
$\mathrm{Pr}(\Psi_1,n)=(1-P_1)\mathrm{Pr}(\Psi_1,n-1)+P_2\mathrm{Pr}(\Psi_2,
n-1)$, and similarly for $\Psi_2$.
These two equations can easily be solved. With a given initial condition we
find
\begin{equation}
\mathrm{Pr}(\Psi_1,n)=\frac{P_2+\lambda^n (P_1\mathrm{Pr}(\Psi_1,0)-P_2
\mathrm{Pr}(\Psi_2,0))}{P_1+P_2},
\end{equation}
with $\lambda=1-P_1-P_2$. For $n\to\infty$ we converge to the equilibrium
probability $\mathrm{Pr}(\Psi_1)=P_2/(P_1+P_2)$. If both distributions have
a finite characteristic waiting time we see that at $n\to\infty$, $\langle
\psi\rangle=(P_2\langle\psi_1\rangle+P_1\langle\psi_2\rangle)/(P_1+P_2)$.

\begin{figure}
\includegraphics[height=7.2cm,angle=270]{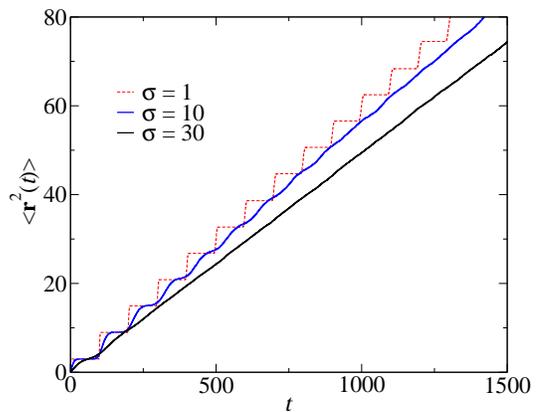} 
\caption{Anticorrelated CTRW for a 3D Gaussian walk. $\psi_1$ and $\psi_2$
are normal distributions centred around $0$ and $100$. The shift times are
$t_1=t_2=50$.}
\label{acorr}
\end{figure}

Consider two limits: (i) if $P_1=1$ and $P_2=1$ (i.e., $t_1\to\infty$ and
$t_2=0$) we change the waiting time at
each step. Roughly, if we begin with the distribution $\Psi_1$ we expect
$t=n\left(\langle\psi_1\rangle+\langle\psi_2\rangle\right)/2+[1-(-1)^n]\left(
\langle\psi_2\rangle-\langle\psi_1\rangle\right)/4$. If $\langle\psi_1\rangle
\ll\langle\psi_2\rangle$, at the beginning we have a short waiting time
alternating
with a long one. After a long time $t\gg\langle\psi_2\rangle$, however, we
see a smooth curve for the mean squared displacement, see Fig.~\ref{acorr}.
(ii) If $P_1\ll1$, we have a classical CTRW governed by distribution
$\psi_1$ that after a while is somewhat modified by distribution $\psi_2$.
At long times  we find the same result as in case (i).
This model can easily be extended to a wider network of nodes.

\section{Conclusions}

We demonstrated that the elusive coupled, non-renewal CTRW can indeed be
applied to non-Brownian processes. Compared to previous models we believe
that the idea of diffusion in the space of waiting times or jump length
is quite intuitive and generic, such that this model will lend itself to
a broad class of phenomena. In particular, this approach allows us to
consider L{\'e}vy stable correlations in waiting times and jump lengths,
significantly generalising previous results. We find that correlated waiting
times lead to subdiffusion while correlations in the jump lengths give rise to
superdiffusion, even when the waiting time or jump length increments are
Gaussian. We also showed that anticorrelations in the long time limit
produce normal diffusion.

Both temporal and spatial correlations are demonstrated to lead to weak
ergodicity breaking: the long time average of the mean squared displacement
of a single trajectory shows different scaling behaviour than the corresponding
ensemble average. Surprisingly this is also the case for jump length
correlations in which successive jumps are separated by constant waiting
times.

\acknowledgments

We acknowledge funding from the Deutsche Forschungsgemeinschaft.

\begin{appendix}

\section{Explicit derivation of the jump statistics}
\label{appendix}

We here show how the average number of jumps for correlated waiting time, and
the position correlations are calculated.

\subsection{Waiting time correlation}

To pass from Eq.~(\ref{Psn}) to (\ref{Psn2}) we note that
\begin{equation}
\sum_{i=0}^{n-1}\psi_{i}=\sum_{i=0}^{n-1}\sum_{j=0}^{i}\delta\psi_{j}=\sum_{
i=0}^{n-1}(n-i)\delta\psi_i.
\end{equation}
For a stable law we know that the sum $Y$ of independent, identically
distributed random variables $X$ fulfils \cite{hughes}
\begin{equation}
Y=\sum_{k=1}^n a_k X_k\Rightarrow Y(s)=X\left(\left(\sum_{k=1}^na_k^{\alpha}
\right)^{1/\alpha}s\right).
\end{equation}
With $\sum_{i=0}^{n-1}(n-i)^2=\sum_{i=1}^ni^2=n(n+1)(n+2)/6$ we see that
in the Laplace domain
\begin{equation}
\prod_{k=0}^{n-1}\psi_{k}(s)=\delta\psi\left(\sqrt{\frac{n(n+1)(2n+1)}{6}}s
\right),
\end{equation}
leading to Eq.~(\ref{Psn2}). The mean number of steps, $\langle n(s)\rangle
=\sum_{n=0}^{\infty}P(s,n)$is then found to be
\begin{equation}
\langle n(s)\rangle=\frac{1}{s}\sum_{n=1}^{\infty}\delta\psi\left(\sqrt{\frac{
n(n+1)(2n+1)}{6}}s\right).
\end{equation}
Due to its Gaussian nature we know that $\delta\psi(s)=1-\tau s+o(s)$, where
$\tau$ is the mean waiting time for a Gaussian random variable of variance
$\sigma$. As we are only interested in the behaviour at long times, i.e.,
small $s$, we approximate $\delta\psi(s)\sim\exp(-\tau s)$ and derive the
leading contribution of $\langle n(s)\rangle$ around $s=0$. With
$n(n+1)(2n+1)/6\approx n^3/3$,
\begin{equation}
\langle n(s)\rangle\sim\frac{1}{s}\sum_{n=1}^{\infty}\exp\left(-\left[n^3/3
\right]^{1/2}s\tau\right).
\end{equation}
Approximating the sum by an integral, we finally obtain Eq.~(\ref{number}).
For a L{\'e}vy stable statistics of the increments $\psi_i$ 
we use the characteristic function $\delta\psi(\omega)=\exp\left(-c|\omega|^{
\alpha}\right)$ in Fourier space denoted by the frequency $\omega$. We find
\begin{equation}
\prod_{i=0}^{n-1}\psi_{i}(\omega)=\exp\left(-\sum_{i=1}^ni^{\alpha}|c\omega|^{
\alpha}\right),
\end{equation}
from which we obtain the average number of steps
\begin{equation}
\langle n(\omega)\rangle=\sum_{n=0}^{\infty}n\frac{1-\exp\left(-(n+1)^{\alpha}
|c\omega|^{\alpha}\right)}{i\omega}e^{-\sum_{i=1}^n i^{\alpha}|c\omega|
^{\alpha}}.
\end{equation}
After approximation of the harmonic number $\sum_{i=1}^ni^{\alpha}\approx
n^{1+\alpha}/(1+\alpha)$ and turning from sum to integral, we find
after Fourier transform
\begin{equation}
\langle n(t)\rangle\sim\frac{(\alpha+1)^{1/(\alpha+1)}}{2\cos\left(
\alpha\pi/[2(\alpha+1)]\right)}\left(\frac{t}{c}\right)^{\alpha/(\alpha
+1)}.
\end{equation}
In the limit $\alpha=2$ we return to expression (\ref{number}).

\subsection{Jump length correlation}

Assume the Gaussian distribution (\ref{gaussjump}) for the jump displacement
$\mathbf{x}(t)=\mathbf{r}(t)-\mathbf{r}(t-1)$
in the $t$th jump, with initial condition $\mathbf{x}(0)=0$. The incremental
change of the jump lengths is $\delta\mathbf{x}(t)=\mathbf{x}(t)-\mathbf{x}(
t-1)$. Consequently, $\langle\mathbf{x}^2(t)\rangle=\frac{3}{2}\sigma^2t$.
We can then calculate the MSD ($\Delta\mathbf{r}(t)=\mathbf{r}(t)-\mathbf{r}
(0)$),
\begin{eqnarray}
\nonumber
\langle\left[\Delta\mathbf{r}(t)\right]^2\rangle&=&\left\langle\left(
\sum_{i=1}^t\mathbf{x}(i)\right)^2\right\rangle
=\left\langle\left(\sum_{i=1}^t\sum_{j=1}^i\delta\mathbf{x}(j)\right)^2
\right\rangle\\
\nonumber
&&\hspace*{-1.4cm}=\left\langle\left(\sum_{i=1}^t(t+1-i)\delta\mathbf{x}(i)
\right)^2\right\rangle\\
&&\hspace*{-1.4cm}=\sum_{i=1}^t\sum_{j=1}^t(t+1-i)(t+1-j)\langle\delta
\mathbf{x}(i)\cdot\delta\mathbf{x}(j)\rangle
\end{eqnarray}
Due to independence of the increments, $\langle\delta\mathbf{x}(i)\cdot\delta
\mathbf{x}(j)\rangle=\frac{3}{2}\sigma^2\delta_{ij}$, we obtain the exact
relation (\ref{eq:msdjump}).

For the jump correlation $\langle\mathbf{r}(t)\cdot\mathbf{r}(t+\tau)\rangle$
we start with
\begin{equation}
\langle\mathbf{x}(t)\cdot\mathbf{x}(t+\tau)\rangle=\left\langle\mathbf{x}(t)
\cdot\left(\mathbf{x}(t)+\sum_{i=1}^{\tau}\delta\mathbf{x}(t+i)\right)\right
\rangle.
\end{equation}
As $\mathbf{x}(t)$ and $\delta\mathbf{x}(t+i)$ are uncorrelated and of mean 0,
we obtain $\langle\mathbf{x}(t)\cdot\mathbf{x}(t+\tau)\rangle=\frac{3}{2}\sigma
^2t$. This expression allows us to obtain the position correlation:
\begin{equation}
\langle\mathbf{r}(t)\cdot\mathbf{r}(t+\tau)\rangle=\left\langle\mathbf{r}(t)
\cdot\left(\mathbf{r}(t)+\sum_{i=1}^{\tau}\mathbf{x}(t+i)\right)\right\rangle.
\end{equation}
Since $\mathbf{x}(t)$ have zero mean we obtain
\begin{equation}
\langle\mathbf{r}(t)\cdot\mathbf{r}(t+\tau)\rangle=\langle(\mathbf{r}(t))^2
\rangle+\sum_{i=1}^{t}\sum_{j=1}^{\tau}\langle\mathbf{x}(i)\cdot\mathbf{x}(t+j)
\rangle.
\end{equation}
As $\langle\mathbf{x}(i)\cdot\mathbf{x}(t+j)\rangle=\frac{3}{2}\sigma^2i$,
we arrive at Eq.~(\ref{autocorr}).

For a L{\'e}vy distribution of the jump increments the position distribution
of the overall process is stable with index $\alpha$ and scale factor
$c(t)\sim ct^{(1+\alpha)/\alpha}/(1+\alpha)^{1/\alpha}$. If we concentrate
on the $x$ coordinate, the characteristic function becomes
$P(q,t)=\exp\left(-\Big|qc(t)\Big|^{\alpha}\right)$.
While the variance of this law diverges, one can calculate fractional
moments of order $0<\delta<\alpha$ \cite{report}, scaling like
$\langle|x|^{\delta}(t)\rangle\simeq c(t)^{\delta}$.
We therefore find the superdiffusive scaling $x^2\sim t^{2(1+\alpha)/\alpha}$.
In the limit $\alpha\to2$, this reproduces the cubic scaling
$x^2\sim t^3$ found for Gaussian correlations in the jump increments. For
decreasing $\alpha$ the superdiffusion is enhanced, and in the limit of
small $\alpha$ we find the scaling $x^2\sim t^{2/\alpha}$ of a regular
L{\'e}vy flight, i.e., a renewal process with a L{\'e}vy stable jump
length distribution $\lambda(\xi)$ of index $\alpha$.

\end{appendix}


\begin{thebibliography}{99}
\bibitem{REM} We denote the Laplace transform by explicit dependence on the
variable: $f(s)=\mathscr{L}\{f(t)\}=\int_0^{\infty}f(t)\exp(-st)dt$.
\end{thebibliography}
\end{document}